\documentclass[doublecol]{epl2}
\usepackage[T1]{fontenc}
\usepackage[english]{babel}
\usepackage[applemac]{inputenc}
\usepackage{amssymb}
\usepackage{mathtools}
\usepackage{sistyle} 
\SIdecimalsign{,} 
\usepackage{enumerate}
\usepackage{graphicx}
\usepackage{tabularx}
\usepackage{longtable}
\usepackage{microtype}

\newcommand{\WW}[1]{\mathcal{W}_{#1}}
\newcommand{\diag}{\mathop{\mathrm{diag}}}
\newcommand{\ddf}{\,\mathrm{d}}
\newcommand{\df}{\mathrm{d}}
\newcommand{\kB}{k_{\text{B}}}
\newcommand{\fL}{f_{\text{L}}}

\title{Monomer dynamics of a wormlike chain}

\author{J.~T.~Bullerjahn \and S.~Sturm \and L.~Wolff \and K.~Kroy}
\shortauthor{J.~T.~Bullerjahn \etal}

\institute{Institut für Theoretische Physik, Universität Leipzig – PF 100920, 04009 Leipzig, Germany}
\pacs{87.15.H-}{Dynamics of biomolecules}
\pacs{66.30.hk}{Self-diffusion in polymers}
\pacs{87.80.Nj}{Single-molecule techniques}

\abstract{ We derive the stochastic equations of motion for a tracer
  that is tightly attached to a semiflexible polymer and confined or
  agitated by an externally controlled potential. The generalised
  Langevin equation, the power spectrum, and the mean-square
  displacement for the tracer dynamics are explicitly constructed from
  the microscopic equations of motion for a weakly bending wormlike
  chain by a systematic coarse-graining procedure. Our accurate
  analytical expressions should provide a convenient starting point
  for further theoretical developments and for the analysis of various
  single-molecule experiments and of protein shape fluctuations.}

\begin{document}

\maketitle

\section{Introduction}
Many of the tools available to an experimental biophysicist probe
either the fluctuations of semiflexible polymers or their response to
external forces. This includes dynamic light scattering
\cite{SchmidtBarmann1989,FargeMaggs1993},
active and passive microrheology of polymer networks
\cite{GislerWeitz1999} or cells
\cite{SemmrichStorz2007}, magnetic bead twisting cytometry
\cite{CaspiElbaum1998}, DNA relaxation and stretching
\cite{PerkinsQuake1994},
single-molecule force spectroscopy
\cite{MeinersQuake2000} and electron
transfer techniques \cite{YangLuo2003}. From a theoretical point of
view, these methods expose different aspects of the \emph{wormlike
  chain} (WLC) model. It provides a minimal description of
semiflexible polymer physics in terms of an inextensible, thermally
fluctuating elastic beam, and has found broad acceptance on grounds of
its excellent agreement with experimental data.

Deformations of the polymer contour excite a broad spectrum of bending
modes with widely disparate relaxation times, thus resulting in
anomalous, subdiffusive dynamics. In the linear regime, valid for
equilibrium fluctuations or small external forces transverse to the
polymer backbone, the dynamic mean-square displacement of a tagged
(but mechanically unaltered) monomer obeys
$\cramped{\text{MSD}_\bot}(t) \propto \cramped{t^{3/4}}$
\cite{Granek1997,DichtlSackmann1999}.  For fluctuations parallel to
the polymer axis, the additional longitudinal solvent friction induces
tension forces which in turn stiffen the polymer and give rise to a
different scaling behaviour, $\cramped{\text{MSD}_\parallel}(t)
\propto \cramped{t^{7/8}}$ \cite{EveraersJulicher1999}. Tension also
dominates the response to
strong point forces and externally imposed
solvent flows; the resulting equations of motion are highly nonlinear
and can produce a multitude of different dynamical regimes even in the
course of a single experiment
\cite{AjdariJulicher1997,SeifertWintz1996,BrochardWyartBuguin1999,
  ObermayerHallatschek2007a,ObermayerHallatschek2007b}.

In many cases of practical interest, a full evaluation of the dynamics
would be needlessly complicated and wasteful, since experimental
manipulation and data acquisition are strongly localised, say, to an
attached tracer particle, or a tagged monomer, in the following simply
referred to as ``the tracer''. Farther parts of the polymer matter
only insofar as they contribute to the force on the tracer. It can
then be preferable to integrate out the polymeric degrees of freedom
beforehand and subsume them under an effective equation of motion
describing the tracer coordinate only.  Such a reduced description is
for example known for the important special case of a tracer subjected
to an externally prescribed deterministic force protocol
\cite{ObermayerHallatschek2007b}. It cannot, however, easily be
extended to accommodate for the fluctuating forces exerted onto the
tracer by an externally controlled confinement potential. Practical
examples that involve such a potential are provided by various
single-molecule manipulation techniques (think \emph{e.g.}\ of an
actin filament labelled with a gold nanoparticle that is trapped by
optical tweezers). The analysis of high-frequency shape fluctuations
of globular proteins, as measured by electron transfer techniques
\cite{MinLuo2005,YangLuo2003} provides another important
example.  Indeed, the WLC has been proposed as one possible
  model of protein fluctuations, but to date only numerical
  evaluations of the corresponding noise and friction functions are
  available within a mean-field approximation to the WLC
  \cite{DebnathMin2005}. 

In the following, we systematically derive the sought-after reduced
equation of motion for the tracer coordinate $x(t)$, starting from the
WLC in the weakly-bending limit, which is asymptotically exact for
large bending rigidity or strong stretching force. The resulting GLE,
\begin{equation}
\label{eq:gle}
\int_{0}^t \mathrm{d}\tau \, K(t-\tau) \dot{x} (\tau) = F(x,t) +
\Xi(t)\, , 
\end{equation}
is not necessarily linear, as it may include an arbitrary external
potential $U(x,t) = -\int \ddf x' F(x'\negmedspace,t)$. In the
infinite-length limit $L\to\infty$, the GLE~(\ref{eq:gle}) can be
worked out explicitly in terms of the microscopic parameters, which
comprise the length $L$ of the polymer, its bending rigidity $\kappa$,
and tension $f$. It is validated by comparison with numerical
solutions of the exact equations of motion. We moreover give a simple
interpolation formula (\ref{eq:interpolation}) that provides a
universal description for polymer-bound tracer particles in strongly
localised externally controlled potentials and possibly also for the
mentioned protein shape fluctuations. We expect it to become a
valuable and convenient tool for analysing a wealth of experimental
data and for future theoretical developments. By providing a
physically transparent and concise parametrisation of measured tracer
movements, it will moreover be helpful in the mutual comparison of
data obtained with diverse experimental techniques. To exemplify our
approach, we calculate various observables characterising the
time-dependent spatial correlations of the tracer motion, such as its
power spectrum and mean-square displacement in presence of a harmonic
trap.

\section{Langevin description of a WLC}
In the WLC model, a semiflexible polymer is mechanically represented
as an inextensible elastic beam of length $L$ and bending rigidity
$\kappa$. It follows that thermal forces can only induce significant
bending on length scales larger than the {\em persistence length\/}
$\ell_p = \kappa/(\kB T)$ (in 3 dimensions). In the weakly-bending
limit, valid for large persistence length $\ell_p \gg L$ or strong
external stretching force $f\gg \kB T/\ell_p$, the polymer is
essentially straight and can thus be treated in terms of its small
excursions $r_\bot(s,t)$ from the straight-rod ground state. The
elastic bending energy in a given configuration $r_\bot(s,t)$ reads
\cite{Granek1997}
\begin{equation}
  \notag
  \mathcal{H} = \int_0^L \mathrm{d} s \, \bigg[ \frac{\kappa}{2} ( r_\bot'' )^2 
  + \frac{f}{2} ( r_\bot' )^2 \bigg].
\end{equation}
Shape fluctuations of the polymer then obey a Langevin equation
obtained by balancing the corresponding bending forces with friction
and thermal (Gaussian white) noise~\cite{Granek1997},
\begin{subequations}\label{eq:eoms}
\begin{align}
\zeta_{\bot} \dot{r}_\bot
\notag
&= - \frac{\delta \mathcal{H}}{\delta r_\bot} + \xi_\bot \\
\label{eq:transversewlc}
&= - \kappa r''''_{\bot} + f r_{\bot}'' + \xi_\bot \\
\langle \xi_\bot \rangle
& = 0 \\
\langle \xi_{\bot} (s,t) \xi_{\bot} (s'\negthickspace,t') \rangle
& = 2 \zeta_{\bot} \kB T \delta (t - t') \delta (s - s').
\end{align}
\end{subequations}
In the vein of a similar treatment for a Rouse chain monomer
\cite{Panja2010a}, we now introduce a tracer at $s=s_0$ and let it
absorb the external driving force $F(x,t)$,
\begin{subequations}
\begin{align}
\label{eq:forceeq}
\zeta_{\text{tr}} \dot{x} (t) &= F(x,t) - \gamma(t) + \xi_{\text{tr}}(t)\\
\notag
\zeta_\bot r_\bot(s,t) &= - \kappa r_\bot''''(s,t) + f r_\bot''(s,t)\\
\label{eq:constrainedwlc}
& \quad + \gamma(t)\delta(s-s_0) + \xi_\bot(s,t)\\
r_\bot(s_0,t) &\stackrel{\smash{!}}{=} x(t) \label{eq:rigidconstraint}.
\end{align}
\end{subequations}
Here $\zeta_{\text{tr}}$ and $\xi_{\text{tr}}$ denote an optional
friction coefficient and Gaussian white noise source for the tracer,
respectively.  Both may be set equal to zero in the tagged-monomer
case. The (Lagrange) forces $\pm\gamma(t)$ serve to rigidly tie the
tracer to the polymer contour at $s_0$, as required by the constraint
(\ref{eq:rigidconstraint}).
\begin{figure}[t]
\centering
  \includegraphics{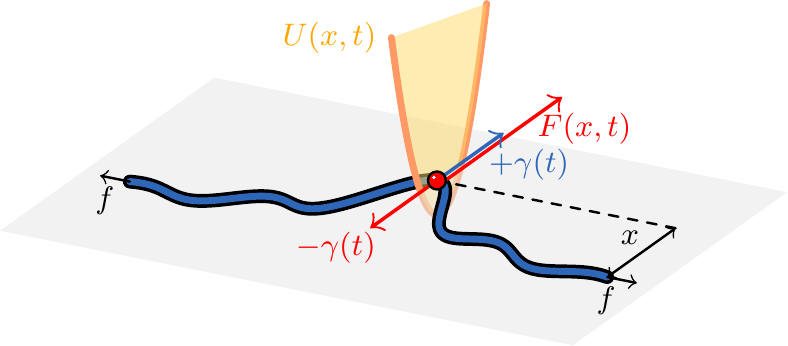}
  \caption{Force diagram for the combined system of a
    tracer (red) and an attached WLC (blue). The tracer is displaced
    by an external potential $U(x,t) = - \int F(x,t)\ddf x$. The
    constraining force pair $\pm\gamma (t)$ fixes the polymer backbone
    to the tracer position~$x$.}
  \label{fig:polymertracer}
\end{figure}
The latter is simplified by transferring
to a comoving reference frame, in which the tracer is always at rest,
$\Delta(s,t) \equiv r_\bot(s,t) - x(t)$. The corresponding equation of
motion acquires a spatially constant friction term due to the drift
between the comoving reference frame and the solvent,
\begin{subequations}
\begin{align}
\notag
\zeta_\bot \dot{\Delta}(s,t) &= -\kappa \Delta''''(s,t) + f \Delta''(s,t) + \xi_\bot(s,t)\\
& \quad + \gamma(t)\delta (s-s_0) - \zeta_\bot \dot{x}(t)\\
\label{eq:transverseDelta}
\Delta(s_0,t) &= 0.
\end{align}
\end{subequations}
The formal solution for $\gamma$ can immediately be written down in
the frequency domain,
\begin{align}
\notag
  \gamma_\omega
  & = - \underbrace{ \frac{\int \mathrm{d}\sigma \, \xi_{\bot,\omega}(\sigma) G_\omega (s_0, \sigma)}{G_\omega (s_0,s_0)}}_{\Xi_\omega}
   - i \omega x_\omega \underbrace{\frac{\int \mathrm{d}\sigma
      \,G_\omega (s_0, \sigma)}{G_\omega
      (s_0,s_0)}}_{-K_\omega} \, ,
\end{align}
where $G(s,s',t)$ denotes the transverse Green's function of a WLC,
\emph{i.e.}~its transverse deformation $r_\bot(s,t)$ in response to a
unit force impulse $\delta(s-s')\delta(t)$. This explicitly
establishes the link between the microscopic equations of motion
(\ref{eq:eoms}) and the coarse-grained equation for the tracer
(\ref{eq:gle}). A practical way of evaluating $\gamma(t)$ numerically
consists in the decomposition of $\Delta(s,t)$ into its normal
coordinates \cite{DoiEdwards1986}. This procedure is discussed for
arbitrary boundary conditions and monomer positions $s_0$
in the appendix.


\section{Analytical solution}
To proceed analytically, we now consider the centre monomer only, $s_0
= 0$. Instead of explicitly including the ``adhesion force''
$\gamma(t)\delta(s)$, which induces a coupling of different eigenmodes
and thus renders the dynamics nondiagonal, we can then easily take
care of the constraint $\Delta(0,t)=0$ by restricting the function
space accordingly. Using only those eigenmodes $\mathcal{W}_n(s)$ satisfying $\mathcal{W}_n(0)=0$, the singular force $\gamma$ can be read off as follows,
\begin{align}
\notag
\kappa \Delta''''(0,t) &= \gamma(t)\delta(s)\\
\notag
\kappa (\Delta'''(0^+,t) - \Delta'''(0^-,t)) &= \gamma(t).
\end{align}
Since the above expression vanishes for odd eigenmodes, only even
modes $\mathcal{W}_n(s) = \mathcal{W}_n(-s)$ contribute to
$\gamma$. Requiring further that $\Delta''''$ constitutes the highest
(and only) singularity, we thus find $\Delta'(0^\pm) = 0$ and so obtain
the following friction kernel,
\begin{equation}
\notag
K(t) = 2 \sum_{n} \mathcal{W}_n'''(0)
\frac{\int_0^{L/2} \mathcal{W}_n(s)\ddf s}{\int_0^{L/2}\mathcal{W}_n(s)^2\ddf s}
e^{-\mathcal{E}_n t/\zeta_\bot},
\end{equation}
where
\begin{align}
\notag
\kappa \mathcal{W}_n''''(s) - f\mathcal{W}_n''(s) &= \mathcal{E}_n \mathcal{W}_n(s)\\
\notag
\mathcal{W}_n(0) = \mathcal{W}_n'(0) &= 0,
\end{align}
and the outer boundary conditions are dictated by the physical
situation. In the infinite-length limit $L\to\infty$, valid for $t\ll
\tau_1$, $K(t)$ becomes independent both of $s_0$ and of the choice of
outer boundary conditions. Using torqued ends $\mathcal{W}_n'(L/2) =
\mathcal{W}_n'''(L/2)=0$ for convenience, we find
\begin{equation}
\label{eq:fullkernel}
K(t) \sim 8\sum_{n=0}^\infty \left[8\frac{n^2\pi^2\kappa}{L^3} + \frac{f}{L}\right]
e^{-t/\tau_n}.
\end{equation}
with a relaxation time spectrum
\begin{equation}
  \notag
  \tau_n \sim \frac{\zeta_\bot L^4}{\kappa \pi^4} \frac{1}{(2n)^4 + (2n+1)^2 f/\fL}.
\end{equation}
\begin{figure}[t]
\centering
  \includegraphics{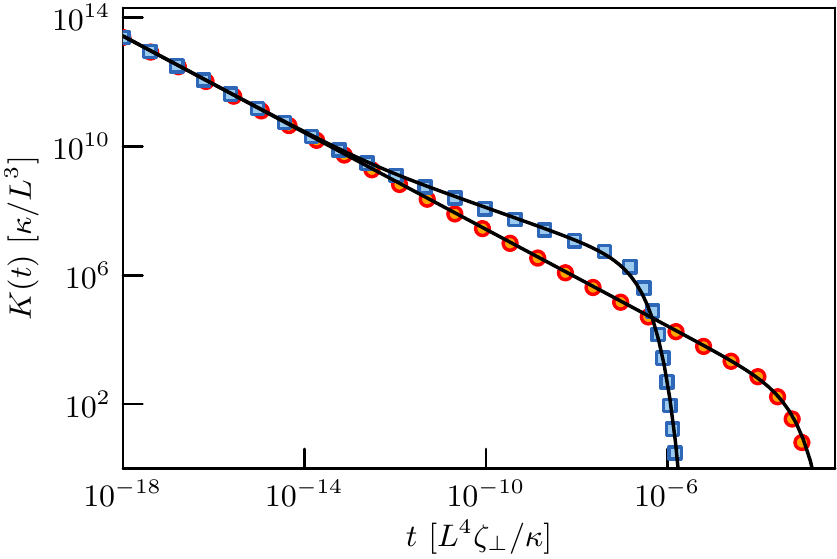}
  \caption{The analytical interpolation formula
    (\ref{eq:interpolation}) compared to the exact numerical solutions
    for $f=0$ ($\bigcirc$) and $f = \cramped{10^6} f_L$~($\square$).}
  \label{fig:kernelcomparison}
\end{figure}
Here, $\fL = \kappa \cramped{\pi^2} / \cramped{L^2}$ denotes the Euler
buckling force. For semiflexible polymers, $\fL$ is usually small in
comparison to externally applied stretching forces.  Mode numbers
below a critical value $n_\text{c} = \cramped{(f/\fL)^{1/2}}$ are then
tension-dominated, whereas shorter-wavelength modes exhibit force-free
relaxation and therefore $\tau_n \propto \cramped{n^{-4}}$. This
divides the frictional response $K(t)$ to a transverse force into
three different asymptotic regimes. At short times $t \ll
\tau_{n_\text{c}} = \zeta_\bot \kappa/ 2 \cramped{f^2}$, the backbone tension
can be neglected so that $K(t)$ simplifies to
\begin{equation}
\label{eq:veryshorttimekernel}
K (t \ll \tau_{n_\text{c}}) \sim \frac{2\sqrt{2} \kappa^{1/4}}{\Gamma (1/4)} \left(\zeta_{\bot}/t\right)^{\mathrlap{3/4}} .
\end{equation}
Between $\tau_{n_\text{c}}$ and the terminal relaxation time $\tau_1$,
we have
\begin{equation}
\label{eq:intermediatetimekernel}
K(\tau_{n_\text{c}} \ll t \ll \tau_{1}) \sim \frac{2}{\Gamma (1/2)} \sqrt{f\zeta_\bot/t} \, ,
\end{equation}
and at very long times $t\gg \tau_1$, the exponential relaxation 
\begin{equation}
\label{eq:longtimekernel}
K (t \gg \tau_{1}) \propto 
\begin{cases}
e^{-t/\tau_1} & f=0\\
e^{-t/\tau_0} & f>0
\end{cases}
\end{equation}
of the lowest mode provides a natural physical cutoff of the
scale-free intermediate asymptotic dynamics.  Using a time-dependent
mode cutoff at $\tau_n = t$, we arrive at the approximate
interpolation formula 
\begin{align}
  \notag
K(t) &\approx \frac{2\sqrt{2}}{3\pi\sqrt{\kappa}}
 \sqrt{\sqrt{f^2 + 4\zeta_{\bot}\kappa/ t} - f} \\
\label{eq:interpolation}
&\qquad\times \left[\sqrt{f^2 +
    4\zeta_{\bot}\kappa/t} + 2f\right]e^{-t/\tau_\ast},
\end{align}
which faithfully reproduces the general solution described in the appendix, see fig.(\ref{fig:kernelcomparison}).

An exact treatment of the binding constraint
(\ref{eq:rigidconstraint}) renormalises the terminal relaxation
time, which is why we consider $\tau_\ast$ as a free (fit)
parameter. In the long-polymer limit ($L,\tau_\ast \to \infty$,
$\cramped{e^{-t/\tau_\ast}} \to 1$), eq.~(\ref{eq:interpolation}) reduces to a
two-parameter formula.

The associated colored noise term $\Xi(t)$ of eq.~(\ref{eq:gle}) can
be determined from the fluctuation-dissipation theorem (FDT),
\begin{equation}
\notag
\langle \Xi (t) \Xi (t') \rangle = 2 \kB T K (\lvert t - t'\rvert).
\end{equation}

\section{Example applications}
As an example application of our single-coordinate equation of motion eq.~(\ref{eq:gle}), we first rederive the time-dependent MSD of a monomer in the bending-dominated regime. We then include an external harmonic potential to compute both the time-dependent MSD and the power spectrum of a polymer-bound tracer particle held in a harmonic trap.

For the free polymer in solution, we set $F(x,t)=0$ and hold the tagged monomer fixed until $t=0$. Using the early-time asymptote to $K(t)$, eq.~(\ref{eq:veryshorttimekernel}), its trajectory then follows as
\begin{equation}
\label{eq:increment}
x(t) - x(0) \sim \frac{1}{2\sqrt{2}\Gamma(3/4)\kappa^{1/4}\zeta_\bot^{3/4}}
\int_{0}^{t} \frac{\Xi (\tau)\ddf \tau}{(t-\tau)^{1/4}} \, ,
\end{equation}
The MSD follows from eq.~(\ref{eq:increment}),
\begin{equation}
  \notag 
  \text{MSD}_{F=0} (t \ll \tau_{n_{\text{c}}}) \sim \frac{1}{2\Gamma (7/4)} \frac{\sqrt{2} \kB T}{\kappa^{1/4}} \left[\frac{t}{\zeta_\bot}\right]^{\mathrlap{3/4}} \, .
\end{equation}
This subdiffusive behaviour coincides with previous theoretical
predictions \cite{FargeMaggs1993} and has been measured directly and
indirectly in networks of polymerized actin
\cite{SchmidtBarmann1989,DichtlSackmann1999}
and microtubuli \cite{CaspiElbaum1998}. Including a stationary
harmonic trap of stiffness $k$, \emph{i.e.}, $F(x,t)\equiv F(x) = -kx$,
equation (\ref{eq:increment}) turns into
\begin{align}
  x(t) \notag
  & \sim x(0) \operatorname{E}_{\frac{3}{4}} \!\bigg[ \frac{-k (t/\zeta_\bot)^{3/4}}{2\sqrt{2} \kappa^{1/4} } \bigg] \\
& \quad - \frac{1}{k} \int_{0}^{t} \mathrm{d} \tau \thinspace \Xi
(\tau) \frac{\partial}{\partial t'} \operatorname{E}_{\frac{3}{4}}\!
\bigg[\frac{-k (t'/\zeta_\bot)^{3/4}}{2\sqrt{2}
  \kappa^{1/4}} \bigg]_{\mathrlap{t'=t-\tau}}\, ,
\end{align}
where
\begin{equation}
\notag
\operatorname{E}_{\alpha} (z) = \sum_{n=0}^{\infty} \frac{z^{\alpha n}}{\Gamma (1 + \alpha n)}
\end{equation}
denotes the Mittag--Leffler function, which can be regarded as a
generalised exponential function. The resulting MSD for $x(0) \equiv
0$ reads
\begin{equation}
\notag
\text{MSD}_{F\neq 0} (t \ll \tau_{n_{\text{c}}}) \sim \frac{\kB T}{k} 
\Bigg[ 1 - \operatorname{E}_{\frac{3}{4}}^{2}\! \bigg[\frac{-k (t/\zeta_\bot)^{3/4}}{2\sqrt{2} \kappa^{1/4}} \bigg] \Bigg] \, .
\end{equation}
\begin{figure}[t]
\centering
  \includegraphics{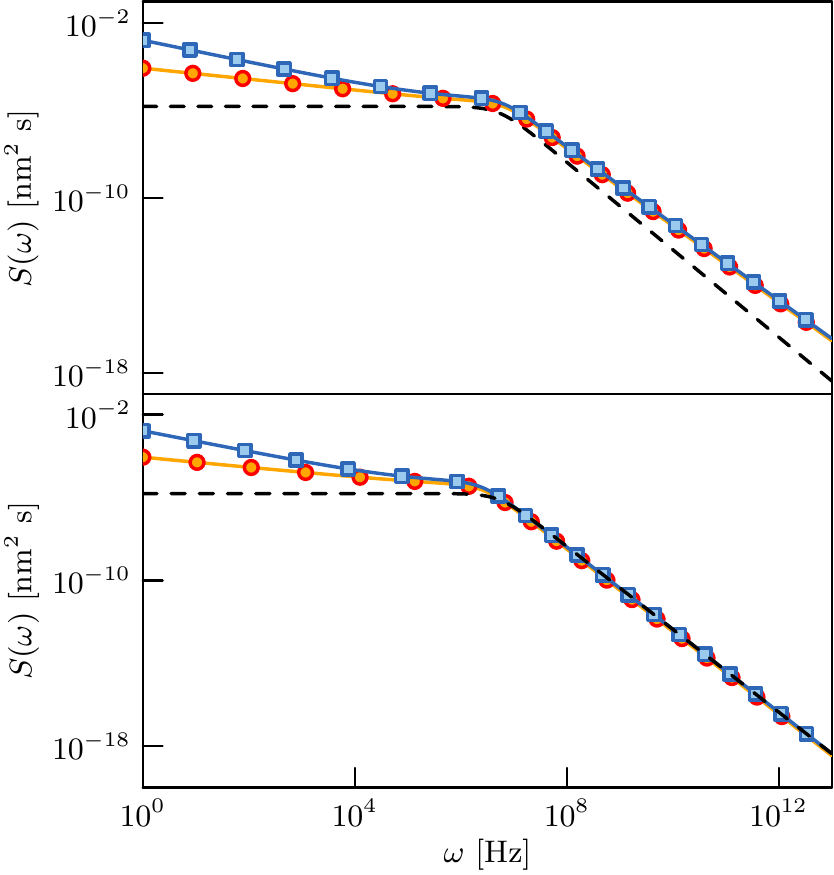}
  \caption{\emph{Top panel}: power spectrum of an unmodified tagged
    monomer (no excess friction) with $\ell_p = \SI{10}{\mu m}$,
    $\zeta_\bot=\SI{10}{mPa\cdot s}$, $k=\SI{1}{pN/nm}$,
    $T=\SI{300}{K}$, $L\to\infty$ (\emph{i.e.}\ $\omega\tau_\ast \to \infty$)
    and $f=\SI{5}{pN}$ ($\square$), $f=0$ ($\bigcirc$). \emph{Bottom
      panel}: power spectrum of an attached tracer particle that has a
    perceptible friction coefficient $\zeta_{\text{tr}}= 6\pi
    \eta_{\text{water}}r$, $r=\SI{10}{nm}$, other parameters as
    above. The Lorentzian power spectrum of the same bead without the
    attached polymer is shown for comparison (dashed line).}
  \label{fig:psdcomparison}
\end{figure}

A deterministic external force $f_{\text{ext}}(t)$ or a dynamically
moving optical trap, represented by $f_{\text{ext}}(t) \equiv k x_0(t)$,
can be included along the same lines by setting $F(x,t) = - k x +
f_{\text{ext}}(t)$. Finally, we calculate the power spectrum density
for a polymer-bound bead in harmonic confinement,
\begin{equation}
  \notag
  S_\omega = \int_{\mathrlap{-\infty}}^{\mathrlap{\infty}}\
  \langle x(t) x(0)\rangle e^{-i\omega t}\df t \;,
\end{equation}
as measured by video microscopy or other in-plane techniques. (Note
that for $3d$ position tracking $x$ becomes a $2d$ vector transverse
to the polymer backbone, and $S_\omega^{\text{3d}}=2 S_\omega$.)

The somewhat lengthy explicit result has a simple structure, suitable
for fitting experimental data. Specialising to the infinite-length
limit ($\omega\tau_\ast \to \infty$), we find 
\begin{align}
 \label{eq:psd}
 \frac{S_\omega}{\kB T} &= \frac{4\zeta_\bot R_\omega +
   2\zeta_{\text{tr}}}{\left[k -
       \omega \zeta_\bot I_\omega\right]^2 + \omega^2
     \left[\zeta_{\text{tr}} + \zeta_\bot
       R_\omega\right]^2}
\end{align}
where $R_\omega$ and $I_\omega$ denote the real and imaginary parts
(with a dimension of length) of
\begin{align*}
   \notag
   2\sqrt{2\kappa} &\bigl[(f
     - \sqrt{f^2 + 4i \omega\zeta_\bot \kappa})^{-1/2}\\
   &\quad + (f +
     \sqrt{f^2 + 4i
       \omega\zeta_\bot \kappa})^{-1/2}\bigr],
\end{align*}
respectively. For a tracer that causes a perceptible friction
($\zeta_{\text{tr}} = 6\pi\eta r_{\text{tr}} > 0$), the ultimate
high-frequency limit is of the usual Lorentzian form $S(\omega \to
\infty) \sim \cramped{\omega^{-2}}$. If $\zeta_{\text{tr}}$ vanishes
or is imperceptibly small compared to the monomer friction, the decay
of the power spectrum is slightly weaker at large frequencies, $S\sim
\cramped{\omega^{-7/4}}$. This can intuitively be understood as a
consequence of the frequency-dependent ``apparent bead size'', given
by the subsection (of length $\ell_\perp$, where $\ell_\perp^{-4}
\simeq \zeta_\perp \omega/\kappa$) of the attached polymer
that equilibrates with the bead within one period $\omega^{-1}$. See
fig.~\ref{fig:psdcomparison} for a graphical representation.

In the absence of backbone tension ($f=0$), the power spectrum
(\ref{eq:psd}) simplifies to ($s = \sin(\pi/8)$, $c=\cos(\pi/8)$)
\begin{align}
  \notag \frac{S^{f=0}_\omega}{\kB T}
  &= \frac{8\zeta_\bot^{3/4}
    (\kappa/\omega)^{1/4}(s+c) +
    2\zeta_{\text{tr}}} {\splitdfrac{\left[k + 2\kappa^{1/4}
        (\zeta_\bot \omega)^{3/4} (s-c)\right]^2}
    {+ \omega^2 \left[\zeta_{\text{tr}} + 2 \zeta_\bot^{3/4}
        (\kappa/\omega)^{1/4}(s + c)\right]^2}}.
\end{align}

\section{Conclusions}
Starting from the formally exact WLC equation of motion in the
weakly-bending rod approximation, we have derived a generalised
Langevin equation describing the dynamics of a tagged monomer of (or
``tracer'' attached to) a semiflexible polymer. The tracer was allowed
to be under the influence of an arbitrary external
potential. Our method is simple, direct and analytically
  solvable. We have furthermore derived a uniformly valid analytic
  interpolation formula which may serve as a compact (two- or
  three-parameter) parametrisation of, \emph{e.g.}, the motion of a
  tracer attached to a semiflexible polymer and manipulated by an
  optical trap, or of conformational fluctuations of protein domains.
With regard to quantitative applications using metallic tracer
particles in combination with optical traps, it might be worthwhile to
extend our results along the lines of \cite{RingsPRL} to take the
heating of the tracer into account.

\acknowledgments We acknowledge financial support from the Deutsche
Forschungsgemeinschaft (DFG) through FOR 877 and the Leipzig School of
Natural Sciences – Building with Molecules and Nano-objects
(BuildMoNa).

\section{Appendix}
For the evaluation of $\gamma(t)$, we decompose $\Delta(s,t)$ into bending eigenmodes $\WW{n} (s)$ and mode amplitudes $a_{n} (t)$, such that
\begin{align}
\notag 
\Delta(s,t) &= \sum_{n=0}^N \WW{n} (s) a_n(t)\\
\notag 
\kappa \WW{n}''''(s) - f \WW{n}''(s) &= \frac{\zeta_\bot}{\tau_n} \WW{n}(s)\\
\notag 
\int_{0}^{L} \mathrm{d}\sigma \, \WW{n} (\sigma) \WW{m}(\sigma) &= \delta_{nm} \, .
\end{align}
The actual physical boundary conditions at the polymer ends dictate the detailed functional form of the $\WW{n}$. A free polymer in solution requires \cite{WigginsRiveline1998}
\begin{equation}
\notag 
\WW{n}''(-L/2) = \WW{n}'''(-L/2) = \WW{n}''(L/2) = \WW{n}'''(L/2) = 0 \, ,
\end{equation}
but the precise choice of boundary conditions is irrelevant to the following discussion. Projecting eq.~(\ref{eq:transverseDelta}) onto each of the $\WW{n}$, we find $N+1$ distinct equations of motion for the $a_n$,
\begin{align}
\notag
\dot{a}_n (t) &= - a_n(t) \biggl( \frac{1}{\tau_n} + \dot{x}(t) \overbrace{\langle \WW{n} (s) \mid 1 \rangle}^{A_n}\\
\tag{A.1}
\label{eq:ampeq}
&\qquad - \underbrace{\frac{\langle \WW{n} (s) \mid \xi_\bot(s,t) \rangle}{\zeta_\bot}}_{\xi_n(t)/\zeta_\bot}
+ \frac{\gamma(t)}{\zeta_\bot} \WW{n}(s_0)\biggr) \, ,
\end{align}
where $\tau_{n}$ is the relaxation time of the $n$th eigenmode.  The singular force term can be eliminated by choosing a complete set of allowed displacements in mode space, $\sum \WW{n}(s_0)a_n(t) = 0$, leading to
\begin{align}
\notag
\delta_1 &= \bigl(\WW{1}(s_0), -\WW{0}(s_0),0,\ldots,0\bigr)\\
\notag
\delta_2 &= \bigl(0,\WW{2}(s_0),-\WW{1}(s_0),0,\ldots,0\bigr)\\
\notag
&\vdots\\
\tag{A.2}
\label{eq:constrainteq}
\delta_{N} &= \bigl( 0,\ldots,0,\WW{N}(s_0),-\WW{N-1}(s_0) \bigr) \, .\notag
\end{align}
Combining the eqs.~(\ref{eq:ampeq}) of motion with the corresponding constraint equations (\ref{eq:constrainteq}), the dynamics of the $a_n$  are determined completely, reading
\begin{equation}
\notag
\label{eq:masterampeq}
M_1 \partial_t (a_0,\ldots,a_n)^\top = M_2(a_0,\ldots,a_N)^\top + \dot{x}(t) V_x + V_\xi (t),
\end{equation}
where $M_{1,2}$ and $V_{x,\xi}$ are given by
\begin{align}
\notag 
M_1 &= \left[\begin{smallmatrix}
\WW{1}(s_0) & -\WW{0}(s_0) &  &  &  \\
 & \WW{2}(s_0) & - \WW{1}(s_0) &  &  \\
 &  & \ddots & \ddots &  \\
 &  &  & \WW{N}(s_0) & -\WW{N-1}(s_0) \\
\WW{0}(s_0) & \WW{1}(s_0) & \cdots & \cdots & \WW{N}(s_0)
\end{smallmatrix}\right]\\
\notag 
M_2 &= \left[\begin{smallmatrix}
-\frac{\WW{1}(s_0)}{\tau_0} & \frac{\WW{0}(s_0)}{\tau_1} & & & \\
& -\frac{\WW{2}(s_0)}{\tau_1} & \frac{\WW{1}(s_0)}{\tau_2} & & \\
& & \ddots & \ddots & \\
& & & -\frac{\WW{N}(s_0)}{\tau_{N-1}} & \frac{\WW{N-1}(s_0)}{\tau_N} \\
0 & 0 & 0 & 0 & 0
\end{smallmatrix}\right]\\
\notag 
V_x &= \begin{bmatrix} A_1\WW{0}(s_0) - \WW{1}(s_0)A_0 \\ \vdots \\ \vdots \\ A_N \WW{N-1}(s_0) - \WW{N}(s_0)A_{N-1} \\ 0 \end{bmatrix} \\
\notag 
V_\xi (t) &= \frac{1}{\zeta_\bot} \begin{bmatrix} \xi_0 (t) \WW{1}(s_0) - \WW{0}(s_0)\xi_1 (t) \\ \vdots \\ \vdots \\ \xi_{N-1} (t) \WW{N}(s_0) - \WW{N-1}(s_0)\xi_N (t) \\ 0 \end{bmatrix} \, .
\end{align}
The solution to eq.~(\ref{eq:masterampeq}) then reads
\begin{align}
\notag
&(a_0,\ldots,a_N)^{\top}(t) = e^{M_1^{-1}M_2(t-t_0)}(a_0,\ldots,a_N)^{\top}(t_0)\\
\notag 
&\quad + \int_{t_0}^t \upd \tau \, e^{M_1^{-1}M_2(t-\tau)}M_1^{-1} \big( \dot{x}(\tau) V_x + V_\xi(\tau) \big) \, .
\end{align}
Inserting the above solution into the original equation of motion (\ref{eq:ampeq}), we obtain the force of constraint in mode space,
\begin{align}
\notag
&\left(\gamma(t) \delta(s-s_0) \right)_n\\
\notag
&\quad = \zeta_\bot \partial_t \dot{a}_n(t) + \zeta_\bot a_n(t)/\tau_n - \xi_n (t) + \dot{x}(t) A_n \zeta_\bot \\
\notag
&\quad = \zeta_\bot \int_{t_{0}}^t \upd \tau \, \big( M_1^{-1}M_2 + \diag(\tau_n^{-1}) \big) \operatorname{e}^{M_1^{-1}M_2(t-\tau)}\\
\notag
&\qquad \times M_1^{-1} \big( \dot{x}(\tau)V_x + V_\xi(\tau) \big)\notag\\
&\qquad\quad + \zeta_\bot \dot{x}(t) (M_1^{-1}V_x + A_n ) \notag \\
\tag{A.3}
\label{eq:solution}
&\qquad\quad + \big( \zeta_\bot M_1^{-1} V_\xi(t)-\xi_{n} (t) \big).
\end{align}
The part outside the integral vanishes, except for the $N$th entry, which is an artefact introduced by the mode cutoff and can safely be ignored: having a finite minimum bending mode wavelength implies that even at arbitrarily short times, a finite part of the polymer will be dragged along with the monomer. 
We thus obtain $\gamma(t)$ as follows,
\begin{align}
\gamma(t)
\notag
& = \lim_{N\to\infty} \bigg[ \sum_{n=0}^N \WW{n}(s_0)^2 \bigg]^{\mathrlap{-1}}\ W^\top \!\! \int_{t_{0}}^t \upd \tau \, \big( M_1^{-1}M_2 \\
\notag 
& \quad + \diag(\tau_n^{-1}) \big) e^{M_1^{-1}M_2(t-\tau)}  M_1^{-1} \big(\dot{x}(\tau)V_x + V_\xi(\tau) \big) \, ,
\end{align}
where $W=\zeta_\bot \big( \WW{0}(s_0),\ldots,\WW{N}(s_0) \big)$. Identifying the random and the $\dot{x}$-dependent terms with $\xi (t)$ and $K (t)$, respectively, we find
\begin{align}
\notag 
K(t) &= W^\top \big( M_1^{-1}M_2 + \diag(\tau_{n}^{-1}) \big) e^{M_1^{-1}M_2 t}M_1^{-1} V_x \, , \\
\xi (t)
\notag
& = W^\top \int_{t_{0}}^{t} \mathrm{d} t' \, \big( M_1^{-1}M_2 + \diag(\tau_{n}^{-1}) \big) \\
\notag 
& \quad \times e^{M_1^{-1}M_2(t-t')}M_1^{-1} V_\xi (t) \, .
\end{align}
Note that if $s_0$ coincides with the center of the polymer, antisymmetric modes will not contribute to $\gamma(t)$; the calculation then has to be restricted to the symmetric component $\Delta_{\text{s}}(s,t)$, otherwise $M_1$ would be degenerate. The above procedure trivially extends to an inhomogeneous stationary force profile $f=f(s)$. In that case, the differential operator $\cramped{\kappa \partial_s^4 - f \partial_s^2}$ turns into $\cramped{\kappa \partial_s^4 - f'(s)\partial_s - f(s)\partial_s^2}$, which changes both the eigenmodes $\WW{n} (s)$ and their respective eigenvalues, but the calculation in mode space remains unaffected.

\bibliography{bibliography}
\end{document}